\documentclass[submission,creativecommons]{eptcs}
\providecommand{\event}{PLACES 2026} 

\usepackage{amsmath}
\usepackage{amssymb}
\usepackage{iftex}
\usepackage{hyperref}
\usepackage{cleveref}
\usepackage{algorithm}
\usepackage{algpseudocode}
\usepackage{enumitem}

\ifpdf
  \usepackage{underscore}         
  \usepackage[T1]{fontenc}        
\else
  \usepackage{breakurl}           
\fi

\usepackage{todonotes}
\usepackage{listings}
\usepackage[normalem]{ulem}

\newcommand{\FORGET}[1]{}

\lstdefinelanguage{hfc}{
	basicstyle=\ttfamily\small, 
	frame=none,
	basewidth=0.5em,
	sensitive=true,
	morestring=[b]",
	morecomment=[l]{//},
	morecomment=[n]{/*}{*/},
	commentstyle=\color{teal},
	keywordstyle=\color{blue}\textbf, keywords={def}, otherkeywords={=>,||,\&\&,!,<=,==},
	keywordstyle=[2]\color{red}\textbf, keywords=[2]{if,else,return,send,retsend,exchange,nbr,let,in,spawn},
	keywordstyle=[3]\color{violet}, keywords=[3]{mux,nfold,max,min,fst,snd,and,or},
	keywordstyle=[4]\color{orange}\textbf, keywords=[4]{somewhere,BIS,gossipEver,gossipMax,diameter_election},
	keywordstyle=[5]\color{blue}, keywords=[5]{false,true,infinity,null}
}

\newcommand{\e}{\mathtt{e}}

\newcommand{\xname}{\mathtt{x}}

\newcommand{\anyvalue}{\mathtt{v}}
\newcommand{\anyvaluealt}{\mathtt{w}}
\newcommand{\lvalue}{\ell}

\newcommand{\neigh}{\rightsquigarrow}

\newcommand{\eventS}[0]{E}
\newcommand{\eventId}[0]{\epsilon}

\newcommand{\deviceId}[0]{\delta}

\definecolor{dark-gray}{gray}{0}

\definecolor{mylightgray}{rgb}{0.97,0.97,0.97}
\definecolor{mygreen}{rgb}{0,0.6,0}
\definecolor{mygray}{rgb}{0.5,0.5,0.5}
\definecolor{mymauve}{rgb}{0.58,0,0.82}

\lstdefinelanguage{Protelis}{
    language=Java,
    morecomment=[l]{//},
    morecomment=[s]{/*}{*/},
    keywords={def, share, if, mux, else, let, in, import, self, public, rep, nbr}, 
    keywords=[2]{false, true, e, pi, infinity, NaN, area, sense,minHood,allHoodPlusSelf,anyHoodPlusSelf,nbrRange,sumHood}, 
    keywordstyle=[2]\color{blue},
    keywords=[3]{isRecentEvent,summarize,S,distanceTo,hopDistanceTo,bisGradient}, 
    keywordstyle=[3]\color{blue}\bfseries,
    keywords=[4]{interior,closure,somewhere,reaches,touches,surroundedBy,implies}, 
    keywordstyle=[4]\bfseries,
    keywords=[5]{densityEstimation,countNearby,crowdTracking,warning,dangerousDensity,gossipEver}, 
    keywordstyle=[5]\color{violet}\bfseries,
    keywords=[6]{env, DIAMETER, OVERCROWDED, AT\_RISK, NONE, PI}, 
    keywordstyle=[6]\color{orange}\bfseries,
    frame=no,
    breaklines=true,
    captionpos=b,
    keepspaces=true,
    backgroundcolor=\color{mylightgray},    
    basicstyle=\linespread{0.5}\ttfamily\small,
    keywordstyle=\bfseries\color{red},
    emphstyle={\bfseries\color{blue}},
    stringstyle=\ttfamily\color{purple},
    commentstyle=\ttfamily\color{mygreen}
}

\lstdefinelanguage{scala}{
  keywords={abstract,case,catch,class,def,%
    do,else,extends,false,final,finally,%
    for,if,implicit,import,match,mixin,%
    new,null,object,override,package,%
    private,protected,requires,return,sealed,%
    super,this,throw,trait,true,try,lazy,%
    type,val,var,while,with,yield,forSome},
  otherkeywords={=>,<-,<\%,<:,>:,\#,@},
  sensitive=true,
  morecomment=[l]{//},
  morecomment=[n]{/*}{*/},
  morestring=[b]",
  morestring=[b]',
  morestring=[b]"""
}

\title{Exploiting Aggregate Programming 
in a \\ Multi-Robot Service Prototype}
\author{Giorgio Audrito
\institute{Università di Torino, Turin, Italy}
\email{giorgio.audrito@unito.it}
\and
Andrea Basso
\institute{MITO Technology, Milan, Italy}
\email{andrea.basso@alumni.epfl.ch}
\and
Daniele Bortoluzzi \qquad Ferruccio Damiani \qquad Giordano Scarso \qquad Gianluca Torta
\institute{Università di Torino, Turin, Italy}
\email{\{name.surname\}@unito.it}
}

\def\titlerunning{AP in a Multi-robot Prototype}
\def\authorrunning{G. Audrito et al.}

\begin{document}
\maketitle

\begin{abstract}
Multi-robot systems are becoming increasingly relevant within diverse application domains, such as healthcare, exploration, and rescue missions.
However, building such systems is still a significant challenge, since it adds the complexities of the physical nature of robots and their environments to those inherent in coordinating any distributed (multi-agent) system.
Aggregate Programming (AP) has recently emerged as a promising approach to engineering resilient, distributed systems with proximity-based communication, and is notably supported by practical frameworks. 
In this paper we present a prototype of a multi-robot service system, which adopts AP for the design and implementation of its coordination software.
The prototype has been validated both with simulations, and with tests in a University library.
\end{abstract}

\section{Introduction}

\label{sec:introduction}

Multi-robot systems are becoming increasingly relevant within diverse application domains, ranging from the exploration of dangerous or extreme environments \cite{ranganathan2010coordinating,cordes2011lunares,portugal:2013}, to rescue missions \cite{habib:2018,queralta:2020}, to service applications, e.g., in healthcare for the elderly \cite{benavidez:2015,di2018multi,pascher2019little}.
Building multi-robot systems represents a significant challenge, because it involves solving two distinct, but interrelated problems.
On one hand, the (possibly heterogenous) hardware characteristics of the robots (sensors, actuators, communication) have to be carefully taken into account and exploited; on the other hand, the distributed, multi-agent nature of multi-robot systems requires the deployment of mechanisms for coordination and collective decision making that perform effectively in real, physical environments.

Aggregate Programming (AP) \cite{bpv:aggregate:programming} has emerged as a generalization of various previous approaches to programming ensembles of devices \cite{SpatialIGI2013,vbdacp:ac:survey:jlamp},  applicable to distributed systems deployed in far edge, fog or cloud environments, as well as (simulations of) swarm robotics \cite{DBLP:conf/isola/AudritoDT22,DBLP:conf/acsos/CasadeiAPV23,AACsi23}. A central feature of this programming model is its ability to express complex distributed processes through function composition.
Aggregate computations are performed over distributed networks of (possibly) mobile devices, each capable of asynchronously performing local computations, and of interacting with a neighbourhood by local exchanges of messages (supported by a peculiar communication primitive named \emph{exchange}).
Several characteristics make AP a promising candidate for tackling the problem of building multi-robot systems (see \Cref{sec:AP} for details). First of all, it embeds resilience (e.g., to failure of communications, failure of nodes) as a first-class citizen, through its self-stabilizing operators; Secondly, it simplifies programming networks of (heterogenous) devices with a compositional, macro-programming style; Finally, it has several practical implementations \cite{DBLP:journals/softx/CasadeiVAP22,url:collektive}, including FCPP\cite{DBLP:conf/coordination/AudritoRT22}, a C++ implementation  that supports deployment on embedded systems.

In this paper, we present in detail the design and implementation of a prototype of a multi-robot
service system for helping users in a University library.
The application was chosen because of the growing presence of service robots in libraries \cite{nguyen2020impact,tella2022robots}, as well as because we had the possibility to experiment in one of the libraries of our University.
While a recent paper applies AP to the formation control of swarms of Rover Wave mini-robots \cite{ABCCDFPV25}, it relies on a centralized deployment of AP which implements the computation and communication of each robot as a local process, leaving to the physical robots just the execution of commands and the production of sensor data. To the best of our knowledge, our system is then the first one to deploy an AP algorithm on real robots.
Our main contributions are: (i) we applied AP to implement the coordination logic of a real multi-robot system, exploiting some of its features to ensure resilience to failures and changes, as well as ease of programming;
(ii) we coped with practical issues not directly addressed by AP, such as the possibility that in some situations two or more robots start the execution of the same task, or the need of efficiently interfacing the AP decision program with the robots navigation control, and sensors feedback;
(iii) we validated our implementation both with a realistic simulator (Gazebo), and with a prototype implementation of the system on a team of physical robots.\footnote{The code for the simulations is publicly available at \url{https://github.com/fcpp-experiments/ros2-library-project}.}

The paper is organized as follows. In \Cref{sec:AP}, we describe the eXchange Calculus underpinning AP.
In \Cref{sec:library-service}, we present our case study, namely a multi-robot library service, and then briefly describe the main employed technologies.
\Cref{sec:algorithm} presents in detail the logic of the AP algorithm we have developed for the case study, which is essentially a distributed, resilient Multi-Robot Task Assignment (MRTA).
In \Cref{sec:experiments} we report on further technical details about experimenting with the Gazebo simulator and with real robots, as well as on the test scenarios we explored.
Finally, in \Cref{sec:conclusion} we close the paper, pointing out directions for further research.

\section{Aggregate Programming}\label{sec:AP}


The reference programming language for AP is the recently proposed {\bf eXchange Calculus (XC)}~\cite{DBLP:conf/ecoop/AudritoCDSV22,DBLP:journals/jss/AudritoCDSV24}, implemented by both the \emph{Field Calculus++ (FCPP)} C++ library \cite{a:fcpp,DBLP:conf/coordination/AudritoRT22,DBLP:journals/scp/AudritoT24}, that will be exploited in this paper, and the \emph{Scala Fields (ScaFi)} Scala library \cite{Casadei:PMLDC16,DBLP:journals/softx/CasadeiVAP22,DBLP:conf/acsos/CasadeiAPV23}. 
XC provides mechanisms to express and compose distributed computations, on a level of abstraction that avoids the explicit management of message exchanges, device position and quantity, and so on. In this context, a single program is periodically and asynchronously executed on every device, according to a cyclic schedule of {\em rounds}. In each round, the device gathers sensors' inputs and recently collected messages; evaluates the program; and broadcasts the result to neighbours and (possibly) actuators.
A formal semantics can be provided to XC programs through the notion of \emph{event structure} \cite{lamport:events}, which is a finite set of events $\eventS$ along with an acyclic \emph{neighbouring relation} $\neigh \subseteq \eventS \times \eventS$ modelling message passing. An event corresponds to a \emph{round of activation} of a device.
\begin{figure}[t]
	\centering
	\includegraphics[scale=1]{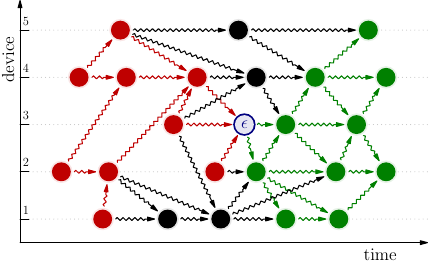}
    \hspace*{0.2in}
	\includegraphics[width=0.5\columnwidth]{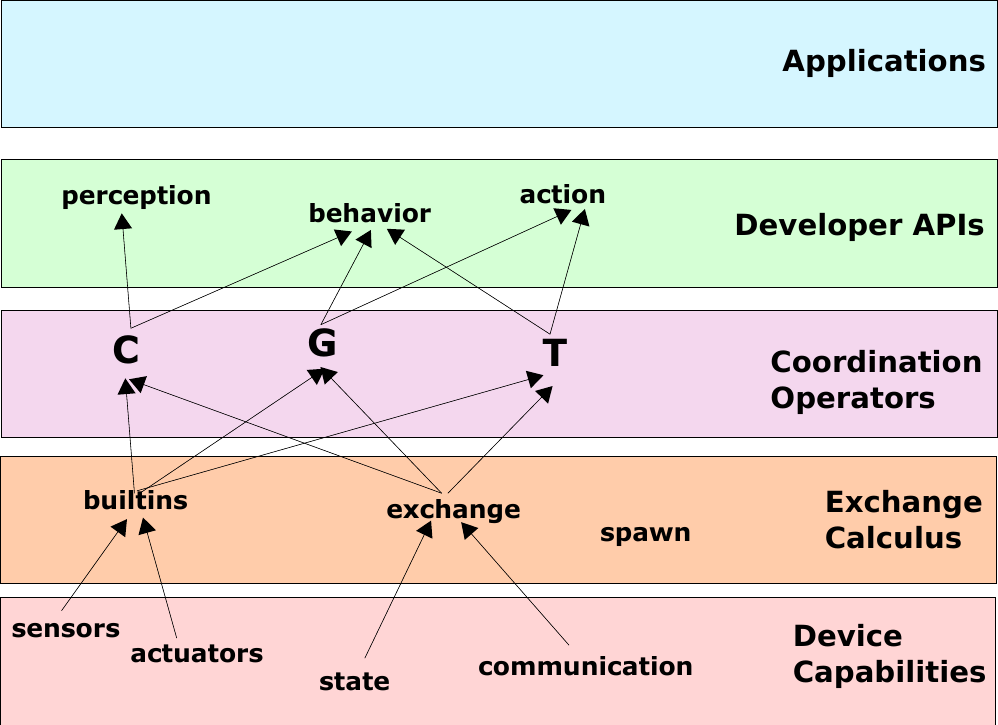}
	\caption{A sample event structure, split in events $\eventId'$ in the past of $\eventId$ ($\eventId' < \eventId$, in red), events in the future ($\eventId < \eventId'$, in green) and concurrent (non-ordered, in black) (left). AP Engineering Stack Layers (adapted from \cite{Viroli-Et-al:COORDINATION-2018}) (right).} \label{fig:structure}
\end{figure}
%

An example structure is illustrated in \Cref{fig:structure} (left). In practice, event structures arise from device neighborhood graphs changing over time. For instance, device $3$ is activated at a certain point in time, with devices $4$ and $1$ as neighbors; after a couple of activation rounds, its neighbors become devices $2$ and $4$.
In XC,  a fundamental role is played by \emph{neighbouring values} $\anyvaluealt$, which represent (sets of) local literals either received from or sent to neighbouring devices. These values are defined as maps $\anyvaluealt = \lvalue[\deviceId_1 \mapsto \lvalue_1, \ldots, \deviceId_n \mapsto \lvalue_n]$, mapping device identifiers $\deviceId_i$ to their respective local literals $\lvalue_i$. Additionally, there is a local literal $\lvalue$ that serves as a \emph{default}, e.g., for neighbours that are unknown to the device, but receive the nvalue. 
The language models local values $\lvalue$ (those not dependent on neighbours) as a special case of neighbouring values $\lvalue[]$, containing only the default local literal. Thus, in XC each value can be considered as a neighbouring value. 
Neighbouring values can be aggregated into a single local value through a functional-style fold operation \lstinline|nfold(f,w,l)|, which applies a specified binary operation \texttt{f} repeatedly to the local values that form \texttt{w}, except for the current device $\deviceId$, whose value is taken from the \texttt{l} argument.

In XC, evaluation is performed within a context that includes all messages received from neighboring devices. A technique known as \emph{alignment} ensures that for each sub-expression, the context is limited to values corresponding to the same sub-expression on neighboring devices. These context values are primarily utilized by a built-in function called {\bf exchange}, which models communication with neighbours. The expression:
\begin{center}
\lstinline[mathescape,basicstyle=\bfseries]{exchange($\e_0$, ($\xname$) => return $\e_r$ send $\e_s$)}
\end{center}
is evaluated according to the following steps.
\begin{itemize}
    \item
    First, gather a {\em neighbouring} value $\anyvaluealt$ that maps each neighbor $\deviceId'$ to the last value shared by $\deviceId'$ for this exchange expression.
    \item
    If it is the first execution of \lstinline{exchange} on the current device $\deviceId$, the value of $\e_0$ is used as the value for $\deviceId$ in $\anyvaluealt$. Otherwise, the value shared by $\deviceId$ itself in its previous round is used instead.
    \item
    Next, evaluate $\e_r$ by substituting $\anyvaluealt$ for $\xname$, resulting in a value $\anyvalue_r$. Do the same for $\e_s$ obtaining value $\anyvalue_s$.
    \item
    Finally, return $\anyvalue_r$ as the value of the \lstinline{exchange} expression in the program, and broadcast $\anyvalue_s$ to neighbors, who will use it in their subsequent rounds to produce their neighboring value $\anyvaluealt$.
\end{itemize}
Consider, as an example, the following function declaration:
\begin{lstlisting}[basicstyle=\footnotesize\ttfamily]
def dist(source) {
  exchange(infinity, (d) => retsend mux(source, 0, nfold(min, d, infinity)+1))}
\end{lstlisting}
In the function declaration above, we use \lstinline|retsend e| as syntactic sugar for \lstinline|return e send e|. Function \lstinline|mux(b,x,y)| is the multiplexer function returning either \texttt{x} or \texttt{y} depending on whether \texttt{b} is true or false, respectively.
The \lstinline/dist/ function calculates hop-count distances from the nearest device where \lstinline/source/ is true, using a single \lstinline/exchange/ construct.

{\bf Aggregate Processes} \cite{DBLP:journals/eaai/CasadeiVAPD21,DBLP:journals/fgcs/AudritoCT24} are distributed computations sustained by the dynamic aggregation of devices. They model transient collective activities, which may concurrently span and overlap over the fabric created by the devices of an AP network.
Specifically, Aggregate Processes thus support the dynamic
injection and execution of collective computations, their diffusion over selected regions of space-time, and their inherent self-adaptation to changes and faults.


For the purposes of this paper, a better intuition on aggregate processes can be gained by understanding a modified version of the \lstinline{dist} function described above, where different parallel computations are spawned by different source devices.\footnote{Note that, in this example, \lstinline{spawn} does not take the additional argument for the process function.} 

\begin{lstlisting}[basicstyle=\footnotesize\ttfamily]
def multi_dist(isSource, theta) {
  val gen_set = if (isSource) {set(uid())} else {set()};
  spawn((i) => {
    val output = dist(uid() == i);
    val status = output <= theta;
    (output, status)
  }, gen_set);
}
\end{lstlisting}

Function \lstinline{multi_dist} spawns a new distance computation process in devices where \lstinline{isSource} is true, limiting the computation to devices within \lstinline{theta} hops from the source.  
Variable \lstinline{gen_set} is a set with the ID of the current device if it is a source, an empty set otherwise.
The \lstinline{spawn} receives \lstinline{gen_set} as the parameter specifying the set of processes it should create (if they don't exist yet). The behavior of the process is as follows: (i) the \lstinline{output} is the value of the \lstinline{dist} function above, computed w.r.t. one source, namely the device that has spawned the process; (ii) the \lstinline{status} is \lstinline{true} (i.e., active) for devices within \lstinline{theta} hops from the source, \lstinline{false} (i.e. terminated) for other devices.


The AP engineering approach goes beyond the features directly built into XC, by defining the {\bf Engineering Stack} shown in Figure \ref{fig:structure} (right).
The first two layers denote the physical \emph{Device Capabilities}, and the \emph{Exchange Calculus} constructs, that provide the first abstraction over such capabilities.
The third layer defines a number of \emph{Coordination Operators} that play a crucial role both in hiding the complexity and in supporting efficient engineering of AP systems. 
Such operators are collective (i.e., computed across the network) and self-stabilising, meaning that when the topology or other network inputs change, after a certain delay also the outputs are updated to their (new) correct values \cite{viroli:selfstabilisation}.
The Figure names three key operators: \emph{G} is
a highly general information spreading and outward computation operation; \emph{C} is its inverse, a general information collection operation; and \emph{T} implements bounded state evolution and short-term memory.
Finally, the top two layers are devoted, respectively, to \emph{Developer APIs} for particular domains, and to \emph{Applications}.

\section{Multi-Robot Library Service}\label{sec:library-service}

The ability to efficiently locate books within a library is crucial for both students and library staff. In 
the University library considered in our study, 
book identification is based on alphanumeric codes, which can be difficult for students to interpret without assistance from librarians. To address this issue, we propose a system that leverages Aggregate Programming to enable a team of robots to collaborate and assist students in locating books.
The use case is structured as follows: (i) A student arrives at the University library and searches for a desired book using an interactive kiosk located at the entrance of the experimental area; (ii) The system processes the query and retrieves the book’s location on the shelves; (iii) The system selects the most suitable robot based on position, battery level, and availability; (iv) The chosen robot navigates to the appropriate shelf and moves toward the book’s precise location.










\begin{figure*}[t]
	\centering
	\includegraphics[width=0.8\linewidth]{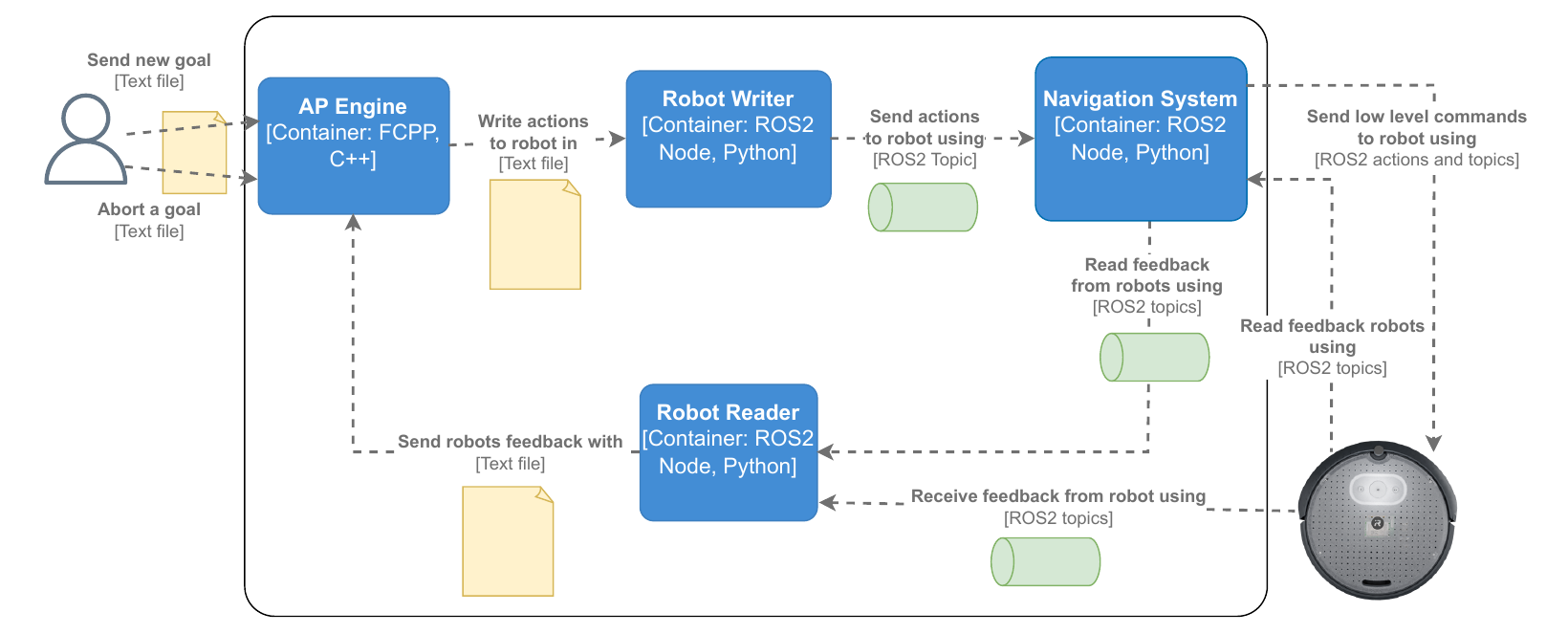}
	\caption{Architecture of the developed platform, integrating Aggregate Programming and ROS2.} \label{fig:architecture}
\end{figure*}

The {\bf system architecture} (\Cref{fig:architecture}) is designed to efficiently coordinate autonomous robots. It integrates a combination of a web-based dashboard, an Aggregate Programming (AP) engine, and ROS2-based robot control components. In this paper, we focus on the AP Engine and ROS2 components.
The process begins with a user searching for a book through the web dashboard. The system retrieves the book’s location among the shelves and generates a text file containing the corresponding coordinates. This file contains the \emph{goal}, and is broadcasted to all AP nodes running on the available robots, either simulated or real.
The \textit{AP Engine} is a layer built on top of the FCPP library, implementing the algorithm described in \Cref{sec:algorithm}, and interfacing with the other system components. The AP nodes communicate to determine the most suitable robot for the task. Once a decision is made, the AP engine generates another text file, containing the assigned \emph{action} to the selected robot.
The ROS2-based \textit{Robot Writer} component interprets the received command and forwards it to the \textit{Navigation System} via ROS2 topics. This component invokes the APIs of the third-party \emph{NAV2} library, which directly interfaces with both the real and simulated robots (also through ROS2) to ensure proper movement toward the shelf coordinates. The \emph{Navigation System} is responsible for handling motion planning and executing potential re-planning when necessary.
Simultaneously, the \textit{Robot Reader} component continuously monitors telemetry data and navigation status through ROS2 topics. These \emph{feedback} files, containing information about localization, status of the task, status of the system, percentage of battery charge, are sent back to the \emph{AP engine}, allowing it to detect failures and, if needed, reassign the task to another robot. 

\section{AP Algorithm}\label{sec:algorithm}


The computational problem that we solve with AP is an instance of Multi-Robot Task Assignment (MRTA), a well known problem that has been widely studied in the literature \cite{CHAKRAA2023104492,GM04}. 
More specifically, according to the taxonomy defined in \cite{GM04}, we address a \emph{ST-SR-IA} variant of MRTA, meaning that robots can execute only a \emph{single task}  at a time (\emph{ST}), that each task can be executed by a \emph{single robot} (\emph{SR}), and that robots get their current task by an \emph{immediate assignment}  (\emph{IA}), instead of a storing a queue of assigned tasks to be executed later.
While the basic \emph{ST-SR-IA} variant can be solved with many existing cooperative (distributed) MRTA algorithms, our proposed solution poses further important requirements that make the adoption of AP compelling: (i) \emph{on-line tasks arrival}: several cooperative assignments should be handled concurrently by the team of robots; (ii) \emph{proximity-based communication}: there should be no need for a centralized communication infrastructure, and the peer-to-peer (P2P) communication may require \emph{multiple hops} to connect each pair of robots; (iii) \emph{adaptivity to failures}: the task assignment should be robust against failures of robots, communication failures, network splits into disconnected sub-networks; (iv) \emph{adaptivity to opportunities}: a robot $R_i$ may opportunistically preempt a task $T_j$ currently executed by another robot $R_k$ if reassigning $T_j$ to $R_i$ is significantly more efficient than the current allocation.
For solving the basic \emph{ST-SR-IA} variant of MRTA, we have adopted a {\em consensus-based} approach \cite{MAHATO2023104270}, where each (idle) robot $R_i$ has a score $r_{i,j}$ for executing task $T_j$, and the team agrees to assign $T_j$ to the robot with the highest score.
Requirement (i) described above is addressed by spawning an aggregate process for each arriving task $T_j$, so that they can be processed concurrently.
Note that, as an alternative to using aggregate processes, we may have each robot explicitly storing and handling a local list of currently active tasks. However, aggregate processes are significantly more convenient, since they transparently ensure that communications about each task assignment are aligned between devices (\Cref{sec:AP}), and handle process propagation and termination.
Requirement (ii) applies to all AP algorithms, given the system model underlying XC, see \Cref{fig:structure} (left). 
The other two requirements (iii),(iv) are mostly addressed by implementing the collective consensus among robots by exploiting one of the leader election operators offered by FCPP. 
Specifically, we consider the variant
\begin{center}
 \lstinline[mathescape,language=hfc]
{diameter_election(value, diameter)}   
\end{center}
which returns the minimum \lstinline{value} among all the devices, given an upper bound to the network diameter (i.e., maximum number of hops between any pair of devices). Since there is eventual consensus about such a minimum, the node holding it can be elected as the \emph{leader}.\footnote{In the context of task assignment, the \lstinline{value} argument should be set to the opposite of the \emph{score} $r_{i,j}$ for assigning $T_j$ to $R_i$, since lower values are preferred.} 
A fundamental feature of such operator for supporting adaptivity, is that it is {\em self-stabilizing} (\Cref{sec:AP}). Specifically, if the current leader disappears (e.g., due to a severe failure), or another node becomes preferred for the leadership (e.g., because the current leader is consuming more energy than normally expected), the operator automatically adapts its output to the changed inputs within a certain delay \cite{MADB22}. 


\Cref{algo:dta-ap} shows the pseudo-code of the program executed in each robot $R_i$, in each round of computation according to the AP paradigm.
Lines 1-4 are the main program, which handles the arrival of a new task $T_j$ to robot $R_i$ by spawning a new {\em Aggregate Process} for it (with $T_j$ as the key of the new process), while also continuing to execute the processes that $R_i$ was already executing.
We don't specify how a new task $T_j$ is sent to robots from the kiosk where the students ask for help, just assuming that it can reach at least one of the robots in the team.
Note that, if more than one robot concurrently creates a new aggregate process for $T_j$, the corresponding collective computations spawned over the network will automatically merge into a single collective computation, since they all share the same key $T_j$.
The behavior of the process is specified in procedure \textsc{TaskProcessing}, which receives the key of the process as its only argument.
First of all, a score $r_{i,j}$ is computed, measuring the quality of assigning $T_j$ to $R_i$, e.g., $r_{i,j}$ should be higher for a robot closer to $T_j$, {\em ceteris paribus} (in \Cref{sec:experiments} we'll define a specific formula).
Score $r_{i,j}$ is then combined into a \emph{pair} $(-r_{i,j}, R_i)$ with the id $R_i$ of the robot, and used as the {\em value} passed to the collective \textsc{LeaderElection} function that returns the minimum value among all connected robots in the network with diameter bounded by $\delta$ (see previous section).
\begin{figure}[t!]
\begin{algorithmic}[1]
    \If{new task $T_j$ arrived to robot $R_i$}
        \State Add $T_j$ to the set of processes to execute
    \EndIf
    \State Execute \textsc{TaskProcessing}($T_k$) for each process active in robot $R_i$
    \vspace{0.5em}
    \Procedure{TaskProcessing}{$T_j$}
        \State $R_i$ computes score $r_{i,j}$ for $T_j$
        \State compute $(r_{best},id_{best})$ with \textsc{LeaderElection}(($-r_{i,j}, R_i)$, $\delta$)
        \If{$R_i$ is leader for $\theta$ consecutive rounds}
            \If{($T_j$ is unassigned) \textbf{or} ($r_{best}$ improves the score by at least $\omega\%$)}
                \State $T_j$ is assigned to $R_i$, which becomes busy
                \State $R_i$ runs \textsc{ExecuteTask}($T_j$) and sets its scores $r_{i,k}$ for $T_k \neq T_j$ to $-\infty$
            \EndIf
        \ElsIf{($R_i$ is executing $T_j$ but another robot is new leader)}
            \State $R_i$ becomes idle and stops executing $T_j$, restoring its scores for $T_k \neq T_j$
        \EndIf
        \If{$R_i$ fails executing $T_j$}
            \State the score $r_{i,j}$ of $R_i$ for $T_j$ becomes constant $-\infty$
        \ElsIf {$R_i$ correctly finishes executing $T_j$}
            \State $R_i$ becomes free and initiates termination of process $T_j$
        \EndIf
        \If {$T_j$ is assigned to  both $R_i$ and $R_k$}
            \State $R_i$ and $R_k$ solve the conflict, so that $T_j$ is assigned to either $R_i$ or $R_k$
        \EndIf
    \EndProcedure
\end{algorithmic}
\caption{Distributed Task Assignment with AP. Parameters $\delta$, $\theta$, and $\omega$ define, respectively, the upper bound on network diameter, the stabilization time for election, and the improvement for re-assignment.}
\label{algo:dta-ap}
\end{figure} 
Scores and the network topology can change, due to robots moving, breaking, etc. but we assume that at some point the election process is stable enough, so that the id $id_{best}$ of the leader is the same for $\theta$ consecutive rounds of execution. 
If the local robot $R_i$ is the potential leader, it becomes so if either $T_j$ is currently unassigned, or the assignment to $R_j$ improves the score by at least $\omega\%$. In such a case, we let $R_i$ to be (at least temporarily) assigned task $T_j$, and starting its execution (lines 10-13), by calling \textsc{ExecuteTask}.
Note that $R_i$ becomes unavailable for other pending tasks $T_k$, with an infinitely low score.
If, on the other hand, the local robot $R_i$ is executing $T_j$ but a better leader is elected, $R_i$ stops executing $T_j$ and becomes available for other tasks (lines 13-15).
%
Lines 16-20 are dedicated to task termination and error handling.
Finally (lines 21-23), although \textsc{LeaderElection} ensures that all the connected robots eventually reach a consensus on the leader, it can happen that two or more robots are assigned the same task when the network of robots becomes (temporarily) disconnected.
When this is detected (because, e.g., the two elected robots get close enough to each other and restore network connection), the {\em conflict} is explicitly resolved by assigning $T_j$ to the  robot with higher score for it.

\section{Validation}\label{sec:experiments}

\subsection{Simulation}

\begin{figure}[t]
	\centering
	\includegraphics[width=0.6\columnwidth]{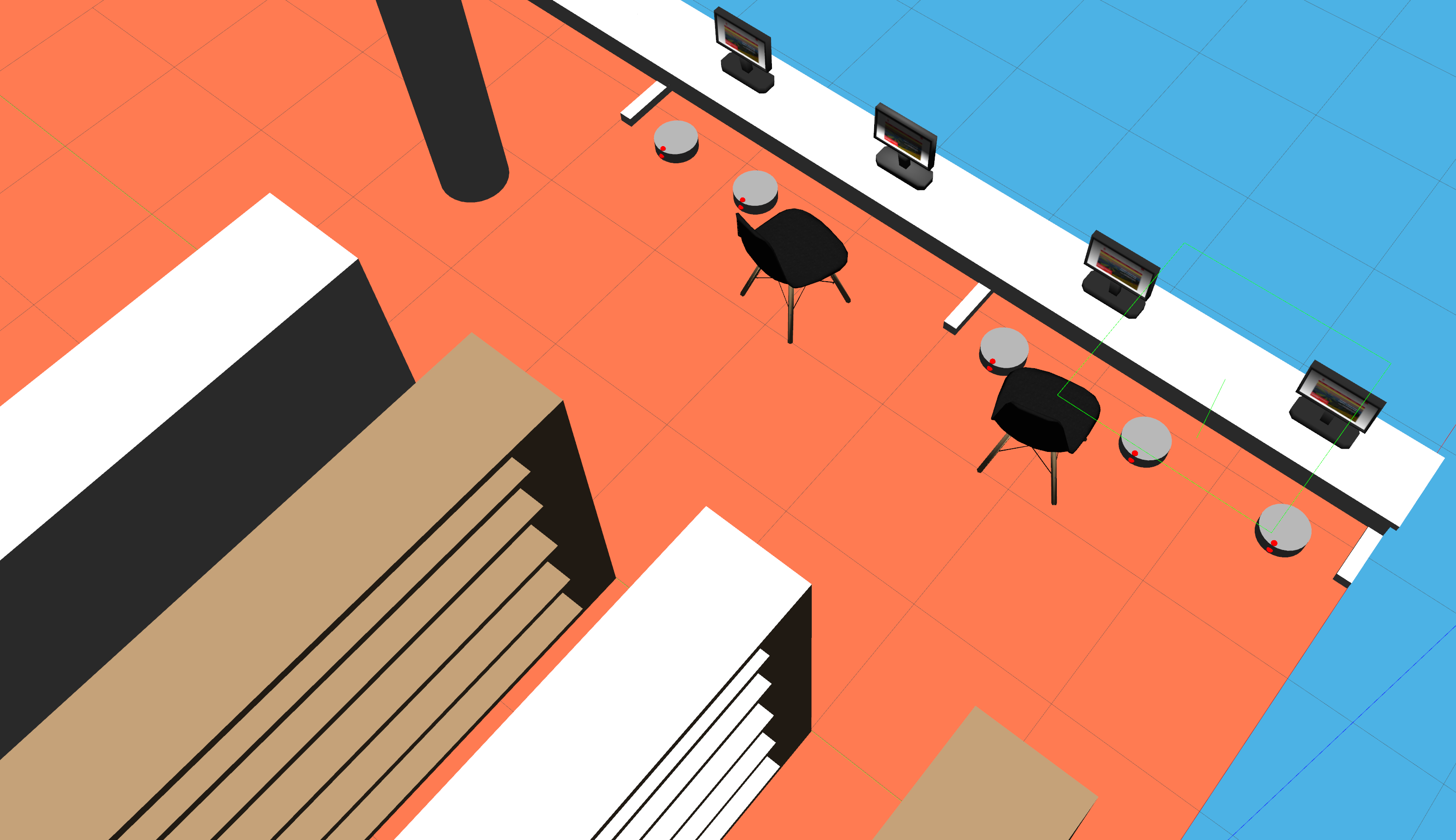}
	\caption{Library world simulated with Gazebo.} \label{fig:simulation}
\end{figure}


\paragraph{Scenario.}

The simulated environment is a near-identical replica of a portion of the 
[removed for double-blind review], as shown in \Cref{fig:simulation}.
In this scenario we used five differential wheeled robots, each equipped with a 360°
2D LIDAR to verify the feasibility of our solution.

\paragraph{Gazebo, FCPP, and ROS2 Configuration.}

Simulation used Gazebo Classic for the physical simulation.
Using SDF we defined the library world with relaxed physical constraints to
increase the number of simulated robots.
To that end we also defined a simplified version of the real robot with simpler
geometry and a lower resolution LIDAR. With these tweaks we were able to simulate
up to ten robots each with its own navigation stack up and running.

The FCPP configuration replicates the real setting in the library, where the FCPP component is configured to operate within a defined area of $3.5\,m \times 6.0\,m$. 
Central to the configuration is the cost function, defined as:
\begin{equation*}
    \text{score} = \text{dist} \times (1.0 - \text{percent\_charge})
\end{equation*}
This function effectively combines the distance to the target with the battery charge level, preferring robots that are both close to the target, and well-powered.
The simulation uses the \emph{lazy} (non-preemptive) version of the algorithm, as described in \Cref{sec:algorithm}. Separately, all user-submitted goals are assigned the same priority, ensuring fair handling during task allocation.
Moreover, the upper bound $\delta$ on the diameter of the network (needed by function \texttt{diameter\_election}, see \Cref{sec:algorithm}) is not statically defined; instead, it is dynamically computed based on the current configuration of the network, in such a way that it is not (too much) larger than the true diameter. 

Since we worked into a multi-robot scenario, we had to use ROS2 \emph{namespaces}
to avoid name collisions in the communication system.
Two topics that required particular attention for remapping were \emph{tf} and \emph{tf\_static}.
In order to integrate with the Aggregate Programming component, the RobotWriter node was in charge of parsing comma-separated values files (CSV) and send the information as a ROS2
message toward the navigation stack.
To produce feedback outside of ROS2, the RobotReader node was in charge of reading information, such as robot's current pose, goal status, battery level, etc., and produce a
CSV file to be fed to the AP Engine.

\paragraph{Test Cases} (the code for simulations can be found at the URL given in \Cref{sec:introduction})
Beyond the default test case, the system supports several variants that enhance its resilience and fault-tolerance by fully leveraging Aggregate Programming.
These behaviors are achieved without explicit reprogramming; instead, they rely on the coordination mechanisms provided by the FCPP library, as explained above in \Cref{sec:algorithm}.
For instance, in the simulation a ROS2 service call can instantly drain a robot's battery, prompting the system to detect the change and reassign the task to an available robot. 
%
\begin{figure}[t]
	\centering
	\includegraphics[width=0.45\columnwidth]{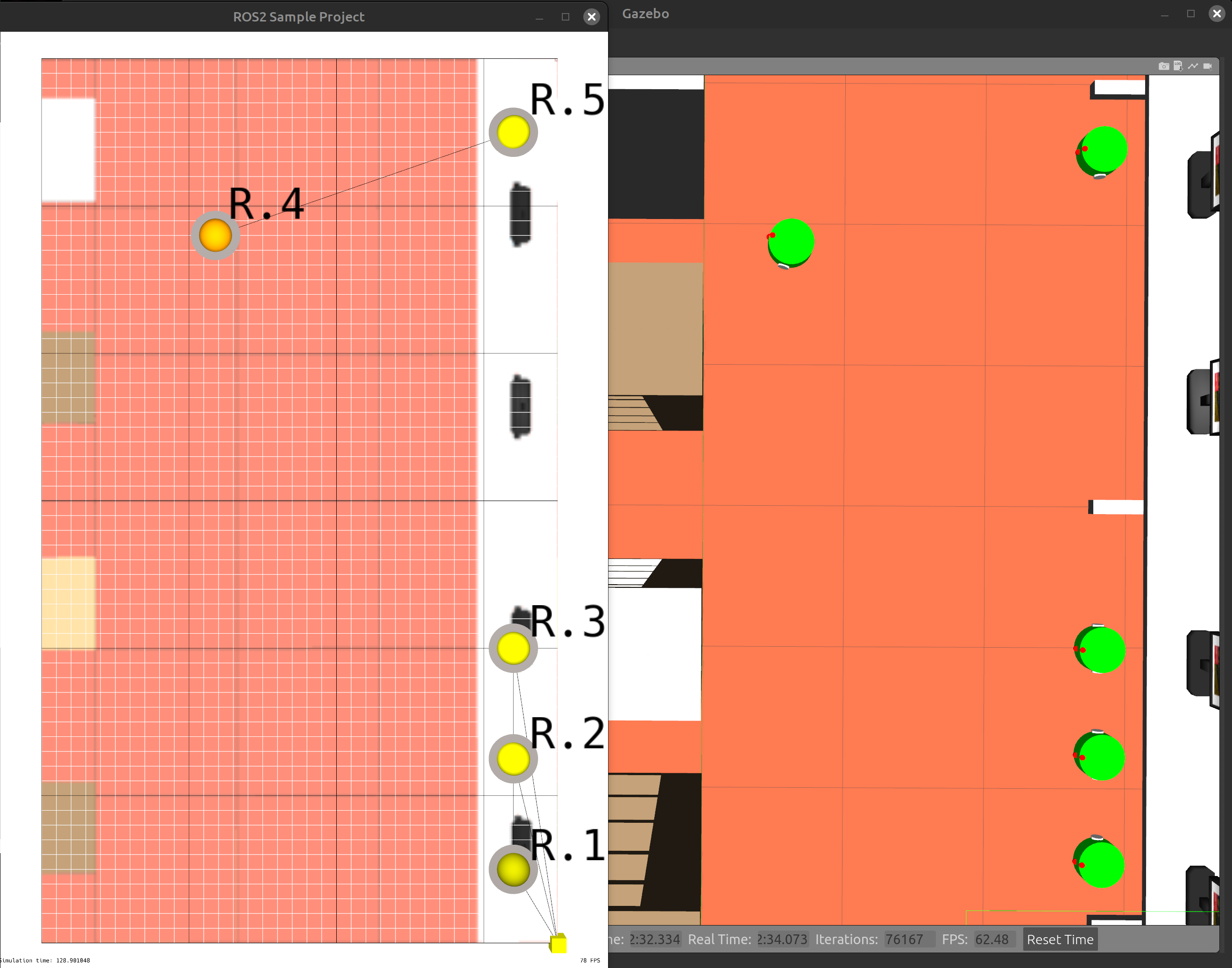}
    \hspace*{0.1in}
	\includegraphics[width=0.45\columnwidth]{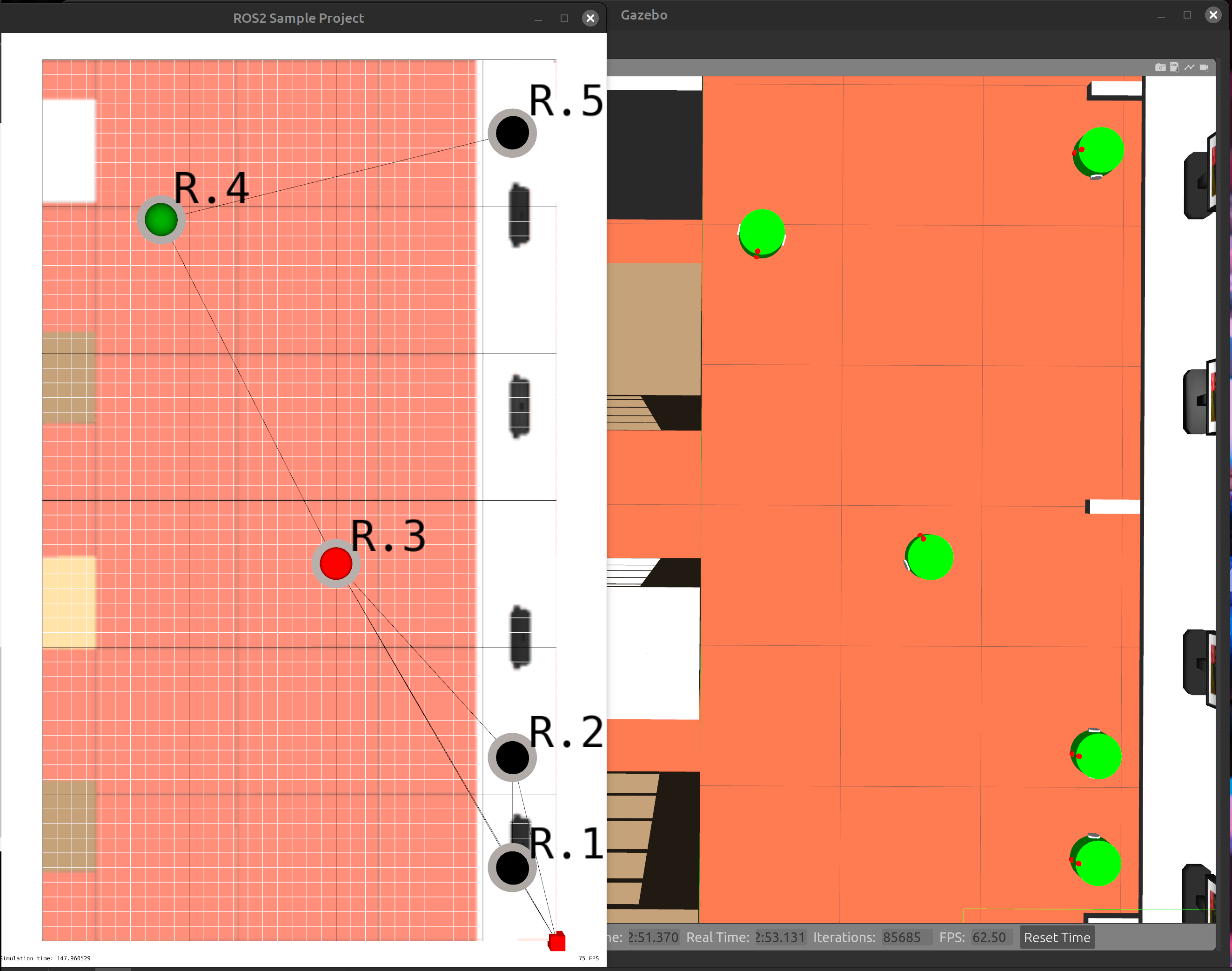}
	\caption{Network partition between robots (left): groups \{1,2,3\} and \{4,5\} (we show both the FCPP simulator, and the Gazebo simulator). Solving the network partition problem (right).} \label{fig:partition}
\end{figure}
Moreover, the Aggregate Programming paradigm ensures resilience to network partitioning in multi-robot systems. To test this use case, the communication range configuration was reduced from 5m to 3m. A network partition occurs when \emph{R.4} moves toward a shelf to execute a task, \Cref{fig:partition} (left). This movement causes \emph{R.4} and \emph{R.5} to become isolated from the rest of the group, as shown by the loss of edges (links between robots are represented as edges between circles). Despite this, tasks can still be initiated independently within each partition.
Once communication is restored, the system detects conflicting task executions, and promptly halts robot \emph{R.3}, marked in red, ensuring consistent operation across the entire network, \Cref{fig:partition} (right).


\subsection{Physical Prototype}

\paragraph{Equipment.}

For the experimentation we used the iRobot Create3 \cite{url:create3} platform as the
base robot extended it with a FHL-LD19P Lidar and a on board MiniPC computer
with 8GB of RAM and a Intel Gemini Lake N4000 processor. 
Power to the computer is provided by the Create3's adapter board, communication
between the robot and the computer is ensured by a Ethernet cable to reduce latency.
Communication between computers for this prototype was realized using a WiFi 6 router.
Another possible solution, closer to the peer-to-peer communication required by AP, would have been to directly setup a WiFi mesh network between the AP nodes. While we were not able to implement it for our tests, this would have just required the definition of a new network driver class for FCPP.


\paragraph{ROS2 and FCPP configuration.}

We chose \emph{CycloneDDS} as the communication middleware, rather than Humble's default DDS (Data Distribution System),
called FastRTPS, because we needed a more configurable system on the network side.
Doing so, we set every robot to communicate only with it's own computer on board and
for every unit, composed by the robot and the added mini-computer, we set a unique ROS domain ID.
A single ROS domain ID is reserved for registering feedback from all the robots, and communication
between the robots is handled using AP's communication scheme.
This setup was necessary because discovery between nodes on the network would overflow the Create3
RAM, causing the robots to continuously restart.
The physical experimentation replicated the configuration used in the FCPP simulator, adapting the communication layer to operate in a real-world network of robots. 
Communication between nodes was handled by a custom UDP-based driver that broadcasts messages in a non-blocking manner over the local wireless network. The driver implements the transceiver interface, and each packet includes the sender's unique identifier, a counter for sequential messages, and the message payload.
Each Aggregate Programming (AP) node (hosted on its associated robot) operated with a round interval of 0.2 seconds, and messages were retained in the AP context for up to 2 seconds to ensure robust inter-robot communication. Moreover, each AP node was assigned a unique ID, configured as a parameter on initialization.
%

\paragraph{Test Cases.}

\begin{figure}[t]
	\centering
	\includegraphics[width=0.4\columnwidth]{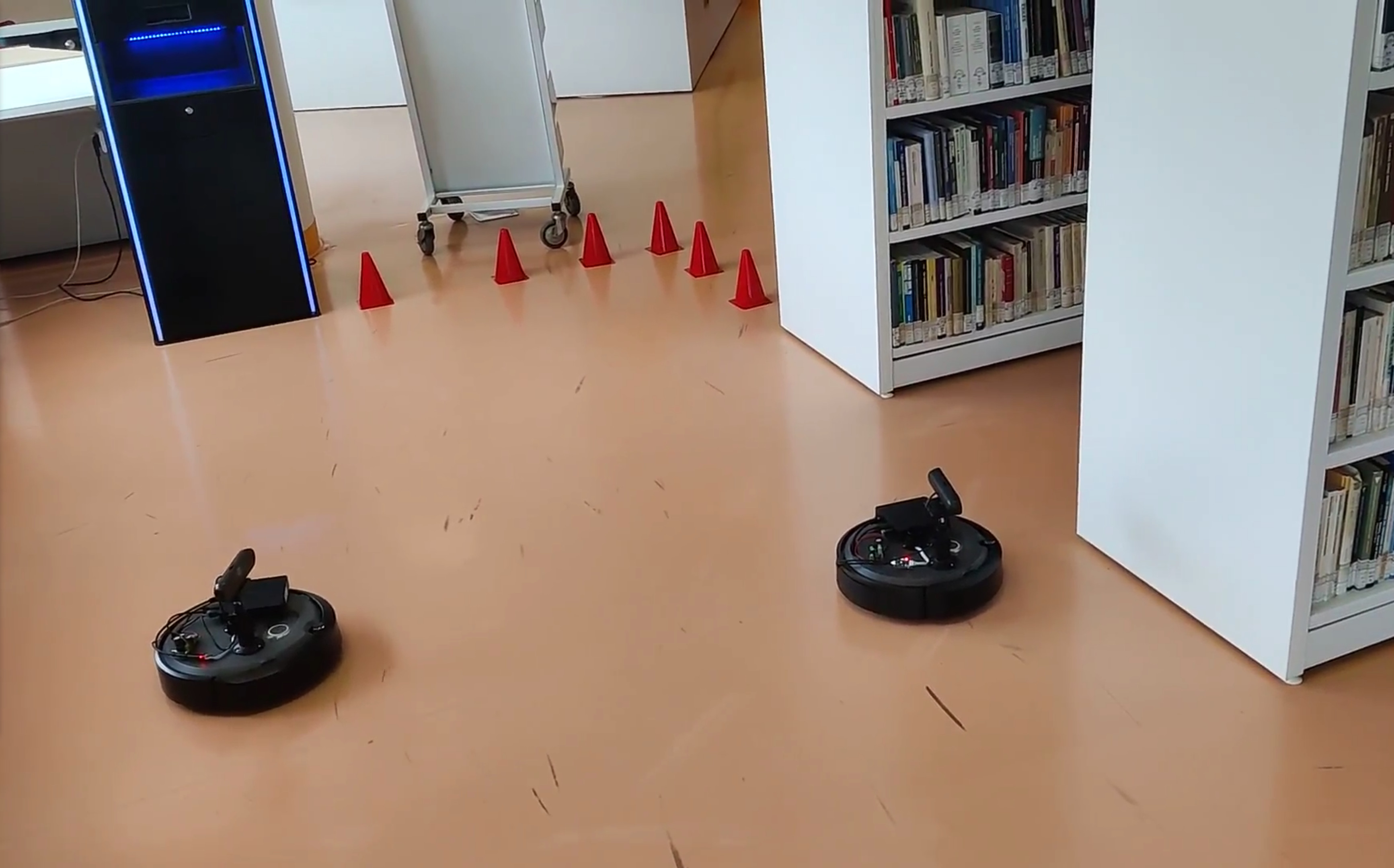}
	\caption{Multiple robots in action.} \label{fig:real-multirobot}
\end{figure}

The same test cases executed in the simulation were also validated on the physical prototype to ensure the system robustness in real-world conditions (\Cref{fig:real-multirobot} shows the real-world deployment using two Create3 robots).
We have implemented a failure notification mechanism that informs other robots whenever a given task cannot be completed due to specific conditions. Examples of such conditions include failures in the navigation stack, the robot being lifted from the ground, critical battery levels, and battery overheating.

\section{Conclusions and Future Work}\label{sec:conclusion}

In this work, we have applied AP to solve a multi-robot coordination problem, with satisfactory results.
First, thanks to the asynchronous, round-based execution of AP, it has been easy to integrate the decision module within the overall architecture including the robot controllers and feedback loops.
Second, the features offered by AP 
have simplified the programming of a quite complex problem, involving the distributed assignment of tasks to robots, dynamic arrival of new tasks, and asynchronous termination/failures of actual execution by the robots.
Last, the self-stabilizing nature and resilience to changes of AP operators, has made it possible to deal naturally with several dynamic/failure scenarios.
%
We have already started working on more complex use cases, e.g., involving the coverage and patrolling of real environments with teams of robots. 
Such scenarios require robots with a more sophisticated collective intelligence, and more hardware (e.g., cameras, outdoor navigation), and processing capabilities (e.g., computer vision). Integrating AP with them is another challenge of our ongoing research.

\bibliographystyle{eptcs}
\bibliography{long}
\end{document}